\pdfoutput=1
\documentclass[11pt,twoside,a4paper,cmspaper,final,collab]{cms-tdr}
\def\svnVersion{2494a24}\def\svnDate{2025/11/22}\def\cmsCernNoTag{CERN-EP-2025-010}\def\cmsCernDate{\today}\def\cmsMessage{Published in the Journal of High Energy Physics as \href{http://dx.doi.org/10.1007/JHEP11(2025)060}{\doi{10.1007/JHEP11(2025)060}.}}
\begin{document}\cmsNoteHeader{HIG-23-010}

\newcommand{\fb}{\unit{fb}}
\newcommand{\cH}{\ensuremath{\PQc\PH}\xspace}
\newcommand{\bH}{\ensuremath{\PQb\PH}\xspace}
\newcommand{\VH}{\ensuremath{\PV\PH}\xspace}
\newcommand{\kappac}{\ensuremath{\kappa_\PQc}\xspace}
\newcommand{\ggH}{\ensuremath{\Pg\Pg\PH}\xspace}
\newcommand{\ttH}{\ensuremath{\ttbar\PH}\xspace}
\newcommand{\DeepJet}{DeepJet\xspace}
\newcommand{\POWHEGBOX} {{\textsc{powhegbox}}\xspace}
\newlength\cmsTabSkip\setlength{\cmsTabSkip}{1ex}
\providecommand{\cmsLeft}{left\xspace}
\providecommand{\cmsRight}{right\xspace}

\cmsNoteHeader{HIG-23-010}

\title{Search for a \texorpdfstring{\cH}{cH} signal in the associated production of at least one charm quark with a Higgs boson in the diphoton decay channel in \texorpdfstring{$\Pp\Pp$}{pp} collisions at \texorpdfstring{$\sqrt{s}=13\TeV$}{sqrt(s) = 13 TeV}}

\date{\today}

\abstract{
This paper presents the first search for a \cH signal sensitive to the coupling of the charm quark (\PQc) to the Higgs boson (\PH) in the associated production of at least one charm quark with a Higgs boson decaying to two photons. The results are based on a data set of proton-proton collisions at a center-of-mass energy of 13\TeV collected with the CMS experiment at the LHC, corresponding to an integrated luminosity of 138\fbinv. Assuming the standard model (SM) rates for all other Higgs boson production processes, the observed (expected) upper limit at 95\% confidence level on the \cH signal strength is 243\,(355) times the SM prediction. Under the same assumption, the observed (expected) allowed interval on the Higgs boson to charm quark coupling modifier, \kappac, is $\abs{\kappac} < 38.1$ ($\abs{\kappac} < 72.5$) at 95\% confidence level.
}

\hypersetup{%
pdfauthor={CMS Collaboration},%
pdftitle={Search for a cH signal in the associated production of at least one chamr quark with a Higgs boson in the diphoton decay channel in pp collisions at sqrt(s) = 13 TeV},%
pdfsubject={CMS},%
pdfkeywords={CMS, Higgs, charm}}

\maketitle 

\section{Introduction}

The Higgs boson (\PH) was discovered~\cite{Aad:2012tfa,Chatrchyan:2012xdj,Chatrchyan:2013lba} by the ATLAS and CMS experiments in 2012, completing the fundamental particle spectrum of the standard model (SM) and setting a milestone for particle physics. Since then, many measurements of the Higgs boson's properties including its spin, mass, and several of its couplings~\cite{ATLAS:2015egz,CMS:2014fzn,CMS:2012vby,ATLAS:2013xga,ATLAS:2016neq,CMS:2017dib,ATLAS:2018pgp,CMS:2020xrn,CMS:2024eka} have been performed. They refine our understanding of the Brout--Englert--Higgs mechanism~\cite{PhysRevLett.13.508,Higgs1964BrokenSM,PhysRevLett.13.321,PhysRev.145.1156} responsible in the SM for the masses of fundamental particles. Any deviation from the SM predictions for the properties of the Higgs boson would provide an indication of possible new physics beyond the SM.

The couplings of the Higgs boson to all third-generation fermions (\PQb quark, \PQt quark, and tau lepton) have been observed~\cite{ATLAS:2014cnc,CMS:2014afl,ATLAS:2014kct,CMS:2013fjq,ATLAS:2014aga,ATLAS:2015muc,CMS:2013zmy,ATLAS:2015xst,CMS:2014wdm,CMS:2017hea,CMS:2018zzl,ATLAS:2022vkf,CMS:2022dwd}. The couplings to second-generation fermions are still under investigation. In the searches for the Higgs boson decay to muons with the CMS and ATLAS experiments~\cite{CMS:2020xwi,ATLAS:2020fzp}, the observed significance at the Higgs boson mass of the incompatibility with
the background-only hypothesis was 3 and 2 standard deviations, respectively. An important next step will be the observation of the Higgs boson coupling to second-generation quarks. Currently, the most sensitive process used to probe Higgs boson couplings to the charm quark, \PQc, is the $\PH \to \ccbar$ decay. The upper limits on the $\PH \to \ccbar$ signal strength based on the full Run 2 data sets of the ATLAS and CMS experiments, are, respectively, 11.5 and 14 times the SM prediction~\cite{ATLAS:vH2025,CMS:2022psv}.
The corresponding observed 95\% \CL interval for the Yukawa coupling modifier of the Higgs boson to charm quarks, \kappac, is $\abs{\kappac} < 4.2$ and $1.1 < \abs{\kappac} < 5.5$, respectively.
Improving the experimental sensitivity to this coupling is one of the major goals of the CMS Higgs boson physics program.
Further, but weaker, constraints could be obtained from searches for a Higgs boson decay to a quarkonium state and a photon~\cite{ATLAS:2022rej,CMS:2018gcm}.

\begin{figure}[!htb]
  \centering
    \includegraphics[width=0.49\textwidth]{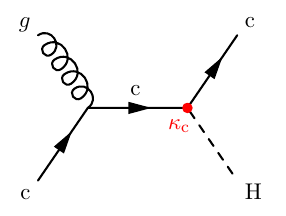}
    \includegraphics[width=0.49\textwidth]{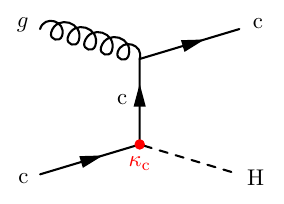}
  \caption{Leading order Feynman diagrams that contribute to the $\Pp\Pp \to \cH$ process. Red dots correspond to vertices where the charm Yukawa coupling modifier \kappac enters.}
  \label{fig:feynmandiagram_cH}
\end{figure}

An additional possibility is to search for a \cH signal containing a $\PQc\PQc\PH$ vertex in the associated production of a Higgs boson and at least one charm quark~\cite{Brivio:2015fxa,Coyle:2019hvs}.
Leading order (LO) Feynman diagrams contributing to the $\Pp\Pp$ (proton-proton) $\to \cH$ process are shown
in Fig.~\ref{fig:feynmandiagram_cH}. These $\kappa_c$-dependent contributions will be indicated with \cH in the following, and correspond to the target signal processes of this analysis.
The ATLAS Collaboration has recently reported on a search for the inclusive production of a Higgs boson and at least one charm quark (the inclusive $\PH{+}\PQc$ process dominated by gluon fusion)~\cite{ATLAS:2024ext}. The ATLAS analysis did not measure the \cH component that is sensitive to the Yukawa coupling of the Higgs boson to charm quarks. Hence no constraint on $\abs{\kappac}$ was reported.

This paper presents the first search for a $\abs{\kappac}$ sensitive \cH signal with a Higgs boson that decays to two photons.
The results are based on the CERN LHC Run 2 data set of $\Pp\Pp$ collisions at a center-of-mass energy of 13 TeV,
collected with the CMS experiment in 2016-2018, corresponding to an integrated luminosity of 138\fbinv~\cite{CMS-LUM-17-003,CMS-PAS-LUM-17-004,CMS-PAS-LUM-18-002}.
The main background processes are the continuous quantum chromodynamics (QCD) diphoton production and \PGg{+}jets processes, where the jet is mis-identified as a photon, and a resonant background from other Higgs boson production mechanisms.
The former does not produce a peak in the diphoton mass spectrum and is estimated from data, while the latter is signal-like and estimated using simulation, assuming the SM hypothesis.
In the SM scenario, \ie, for $\kappac = 1$, \cH production has a very low cross section compared to the main background processes. However, in certain scenarios beyond the SM, $\kappac$ could be significantly different from one and therefore result in a higher \cH production cross section~\cite{Delaunay:2013pja,Perez:2015aoa,Ghosh:2015gpa,Botella:2016krk,Harnik:2012pb,Altmannshofer:2016zrn}.

\section{The CMS detector}
The central feature of the CMS apparatus is a superconducting solenoid of 6\unit{m} internal diameter, providing a magnetic field of 3.8\unit{T}. Within the solenoid volume there are a silicon pixel and strip tracker, a lead tungstate crystal electromagnetic calorimeter (ECAL), and a brass and scintillator hadron calorimeter (HCAL), each composed of a barrel and two endcap sections.
ECAL consists of 75\,848 lead tungstate crystals, which provide coverage in pseudorapidity $\abs{\eta} < 1.48 $ in a barrel region and $1.48 < \abs{\eta} < 3.0$ in two endcap regions (EE). Preshower detectors consisting of two planes of silicon sensors interleaved with a total of 3 radiation length of lead are located in front of each EE detector.
Forward calorimeters extend the $\eta$ coverage provided by the barrel and endcap detectors. Muons are reconstructed in gas-ionization detectors embedded in the steel flux-return yoke outside the solenoid. More detailed descriptions of the CMS detector, together with a definition of the coordinate system used and the relevant kinematic variables, can be found in Refs.~\cite{CMS:2008xjf,CMS:2023gfb}.

Events of interest are selected using a two-tiered trigger system.
The first level, composed of custom hardware processors, uses information from the calorimeters and muon detectors to select events at a rate of around 100\unit{kHz} within a fixed latency of about 4\mus~\cite{CMS:2020cmk}.
The second level, known as the high-level trigger (HLT), consists of a farm of processors running a version of the full event reconstruction software optimized for fast processing, and reduces the event rate to around 1\unit{kHz} before data storage~\cite{Khachatryan:2016bia,CMS:2024aqx}.

\section{Simulated samples}

Signal and background processes are simulated with several Monte Carlo (MC) programs.
The generator \POWHEGBOX~\cite{Nason:2004rx,Frixione:2007vw,Alioli:2010xd} at next-to-leading order (NLO) in QCD is used to simulate Higgs boson production through gluon-gluon fusion (\ggH)~\cite{Bagnaschi:2011tu}, while the generator \MGvATNLO~\cite{Alwall:2007fs,Frederix:2012ps,alwall2014automated} (version 2.4.2) at NLO in QCD is used to generate Higgs boson production processes through vector boson fusion (VBF), associated production with a vector boson (\VH, $\PV=\PW$/\PZ),  associated production with a pair of top quarks ($\ttbar\PH$), associated production with a bottom quark (\bH), and associated production with a charm quark (\cH). The process where a Higgs boson is produced in association with a single top quark ($\PQt\PH$) is not included since its contribution is negligible.
The cross sections and branching fractions for the processes are taken from the LHC Higgs Cross Section Working Group report~\cite{deFlorian:2016spz}.

For the \cH signal and the \bH background processes, a new dedicated simulation is designed specifically for this analysis.
This simulation considers the most important production modes, which are proportional to the square of the charm and bottom Yukawa couplings for the \cH and \bH processes, respectively.
The contributions coming from the interference between \cH/\bH and \ggH terms and the \PH{}+\PQc and \PH{}+\PQb parts that are not sensitive to the Yukawa couplings of the \PQc and \PQb quarks were found to be negligible for the sensitivity reached by this analysis.
In these associated production processes there are two ways of performing the fixed order amplitude calculation, i.e. depending on whether the \PQc and \PQb quarks are considered to be massive or not in the computation.
By convention, \PQb quarks are considered massive in the 4-flavor scheme (4FS) and massless in the 5-flavor scheme (5FS).
This applies in the same way to the \PQc quarks in the 3- and 4-flavor schemes, respectively.
These approaches are supposed to address issues that arise in different kinematic regimes, which one may classify by considering if the mass of the quark plays the role of a hard scale in the process under investigation.
Considering the quarks as massless objects at the level of short-distance interactions has the advantage of resulting in much simpler calculations than their massive counterparts, but leads to divergences at low energies so that such events are not simulated down to $\pt = 0$ at matrix element level.
When matched to parton showers (PS) though, this information can be recovered thanks to the backward evolution of the initial-state \PQc and \PQb quarks.
This will generate heavy-flavor hadrons in the final state, thus recovering realistic \PQc- and \PQb-tagging requirements, regardless of their transverse momenta~\cite{bbH:Wiesemann_2015}.

After studying the dependence of the cross section and the kinematics with respect to the FS, as well as the treatment of the renormalization ($\mu_R$), factorization ($\mu_F$) and PS scales in the simulation, by comparing samples with different starting values for these parameters and exploring both a fixed and dynamic scale choices~\cite{alwall2014automated}, the nominal \cH and \bH samples used in the analysis were generated treating the charm and bottom quarks as massless (4FS for \cH and 5FS for \bH).
A dynamic treatment of the renormalization and factorization scale parameters was adopted, each with a starting value of $\HT/4$, where $\HT=\sum_i\sqrt{\smash[b]{m_i^2+p_{Ti}^2}}$ and $i$ runs over the outgoing parton of the fixed order calculation.
This massless-quark approach enables the use of the FxFx merging procedure~\cite{Frederix:2012ps} within the NLO simulation, thus improving the description of the kinematic observables by reducing the dependence on the PS for additional emissions.
The cross sections are found to be $\sigma_{\cH} = 90\fb$ and $\sigma_{\bH} = 660\fb$, where the calculation is obtained from a dedicated study following the method presented in Ref.~\cite{bbH:Wiesemann_2015}.
An uncertainty covering the residual discrepancies in the yield prediction between massive and massless FSs after the scale optimization, mostly due to cross section differences and the distribution of the \PQc-quark jet $\eta$, was extracted by comparing the expected event counts obtained using a massive and a massless FS of each process (\cH, \bH). The typical size of this uncertainty is 25--30\%. 

The signal and resonant background samples are processed with \PYTHIA version 8.240~\cite{Sjostrand:2014zea} for modeling of parton showering, hadronization and underlying events with the CUETP8M1~\cite{Khachatryan:2015pea} (CP5~\cite{Sirunyan:2019dfx}) tune for the simulation of 2016 (2017--2018) data.
Parton distribution functions (PDFs) are taken from the NNPDF 3.0~\cite{NNPDF3} (3.1~\cite{NNPDF31}) set, when simulating 2016 (2017--2018) data.
In contrast, the diphoton background ($\PGg\PGg${+}jets) is generated with \SHERPA version 2.2.4~\cite{10.21468/SciPostPhys.7.3.034}, which produces tree-level processes with up to three additional jets, as well as loop-induced processes.
The single-photon (\PGg{+}jet) background is modeled with \PYTHIA~8.212 at LO.
The continuous background MC samples are used only for the optimization of the analysis, while the background model for signal extraction is entirely derived from the data.
For all the processes, \GEANTfour~\cite{Agostinelli:2002hh} is used to simulate the detector response.
The simulated events include additional $\Pp\Pp$ interactions within the same or nearby bunch crossings (pileup), as observed in the data.

\section{Object selection}

The primary collision vertex is identified using a multivariate technique based on a boosted decision tree (BDT)~\cite{Sirunyan:2018ouh}. Charged hadrons originating from other vertices are removed from the analysis.

A global ``particle-flow'' algorithm~\cite{CMS:2017yfk} aims to reconstruct all individual particles in an event, combining information provided by the all-silicon inner tracker and by the ECAL and HCAL, with data from the gas-ionization muon detectors.
Photons are identified as ECAL energy clusters not linked to the extrapolation of any charged-particle trajectory to the ECAL.
Electrons are identified as a primary charged-particle track and potentially many ECAL energy clusters corresponding to this track extrapolation to the ECAL and to possible bremsstrahlung photons emitted along the way through the tracker material.
Muons are identified as tracks in the central tracker consistent with either a track or several hits in the muon system, and associated with calorimeter deposits compatible with the muon hypothesis.
Charged hadrons are identified as charged-particle tracks neither identified as electrons, nor as muons, also using HCAL information.
Finally, neutral hadrons are identified as HCAL energy clusters not linked to any charged-hadron trajectory, or as a combined ECAL and HCAL energy excess with respect to the expected charged-hadron energy deposit.

The photon energy is measured with the ECAL. The energy of electrons is determined from a combination of the track momentum at the main interaction vertex, the corresponding ECAL cluster energy, and the energy sum of all bremsstrahlung photons attached to the track. The energy of muons is obtained from the corresponding track momentum. The energy of charged hadrons is determined from a combination of the track momentum and the corresponding ECAL and HCAL energies, corrected for the response function of the calorimeters to hadronic showers. Finally, the energy of neutral hadrons is obtained from the corresponding corrected ECAL and HCAL energies.

For each event, hadronic jets are clustered from these reconstructed particles using the infrared- and collinear-safe anti-\kt algorithm~\cite{Cacciari:2008gp, Cacciari:2011ma} with a distance parameter of 0.4. Jet momentum is determined as the vector sum of all particle momenta in the jet, and is found from simulation to be, on average, within 5-10\% of the true momentum over the whole \pt spectrum and detector acceptance. Pileup interactions can contribute with additional tracks and calorimetric energy depositions to the jet momentum. To mitigate this effect, charged particles identified to be originating from pileup vertices are discarded and an offset correction is applied to correct for the remaining contributions~\cite{Sirunyan:2020foa}.
Jet energy corrections derived from simulation are used to bring the measured jet energy to that of particle-level jets on average. In situ measurements of the momentum balance in dijet, $\text{photon}+\text{jet}$, $\PZ+\text{jet}$, and multijet events are used to account for any residual differences in the jet energy scale between data and simulation~\cite{CMS:2016lmd}. The jet energy resolution is 15--20\% at 30\GeV, 10\% at 100\GeV, and 5\% at 1\TeV~\cite{CMS:2016lmd}. Additional selection criteria are applied to each jet to remove jets potentially dominated by anomalous contributions from various subdetector components or reconstruction failures~\cite{Sirunyan:2020foa}.

The photons used in the analysis are required to satisfy an identification (photon ID) criterion based on a BDT classifier trained to separate photons from jets~\cite{CMS:2020uim}.
Inputs to the photon ID BDT, such as shower shape and isolation variables in the simulation, are corrected with a chained quantile regression method~\cite{DBLP:journals/corr/abs-1211-6581} to match the data better than a few percent across the whole spectrum of the output BDT.

Jets originating from charm quarks are identified by a \DeepJet algorithm based on deep neural networks~\cite{Bols:2020bkb,CMS-DP-2018-058,Sirunyan:2017ezt}.
For each jet, the algorithm outputs a CvsL score to separate charm from light-quark or gluon (\PQu, \PQd, \PQs, \Pg) jets.
The chosen working point has a typical efficiency for charm jets of $\sim$30\% and a typical rejection for light-quark or gluon jets of $\sim$95\%.
The \DeepJet algorithm also provides a CvsB score to discriminate charm and bottom quark jets.
However, given that the dominant backgrounds are the continuous background and the \ggH process (most events of which do not contain bottom quark jets),
the CvsB score provides only minor improvement in analysis sensitivity and hence is not used.

\section{Event selection}
\label{sec:selection}

Events are selected at the HLT using a diphoton trigger level~\cite{CMS:2024aqx} that includes a calorimeter-based photon identification requirement, with photon transverse momentum ($\pt^{\PGg}$) thresholds of 30 and 18\GeV for the leading and subleading photons, respectively, in 2016 data. For the 2017--2018 data-taking periods, the subleading $\pt^{\PGg}$ threshold is increased to 22\GeV. 

To target the $\PH \to \PGg\PGg$ decay, events are required to have two identified photon candidates that are within the ECAL and tracker fiducial region ($\abs{\eta}<2.5$), excluding the ECAL barrel-endcap transition region ($1.44<\abs{\eta}<1.57$).
The diphoton candidates are required to pass the following criteria: $100 < m_{\PGg\PGg} < 180\GeV$,
$\pt^{\PGg 1}/m_{\PGg\PGg} > 1/3$, and $\pt^{\PGg 2}/m_{\PGg\PGg}> 1/4$, where $m_{\PGg\PGg}$ denotes the diphoton mass and $\pt^{\PGg 1}$ ($\pt^{\PGg 2}$) denotes the transverse momentum of the leading photon (subleading photon).
The diphoton mass resolution, as measured in $\PH \to \PGg\PGg$ decays, is typically in the 1--2\% range, depending on the measurement of the photon energies in the ECAL and the event topology~\cite{CMS:2020xrn}.
These criteria were optimized in the context of the inclusive $\PH \to \PGg\PGg$ analysis~\cite{CMS:2020xrn} and are suitable for the \cH analysis as well.

\begin{figure}[!htb]
  \centering
  \includegraphics[width=0.7\textwidth]{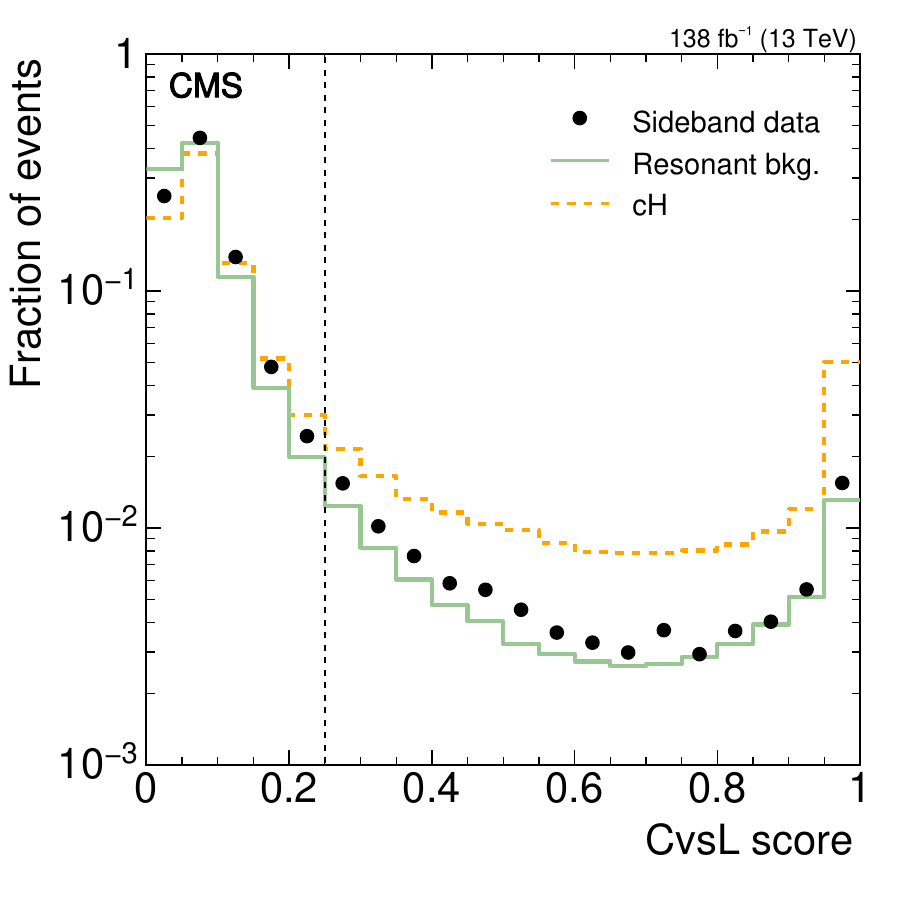}
  \caption{Distributions of the CvsL score for the leading jet in simulated \cH signal and resonant background events and sideband data events. Error bars representing the statistical uncertainties in the data are too small to be displayed. Events with CvsL score values below the dashed vertical line are not included in the signal region.}
  \label{CvsLandCvsB}
\end{figure}

Furthermore, to target the associated charm quark in the signal process, events are required to have at least one jet with $\pt > 25\GeV$ and $\abs{\eta}<2.5$.
The jet with the largest \pt must be \PQc tagged by requiring a CvsL score $>$0.25.
The CvsL score distributions for the leading jet in simulated \cH signal and resonant background events, as well as sideband data events, are shown in Fig.~\ref{CvsLandCvsB}.
Sideband data events are selected in the diphoton signal mass intervals $100 < m_{\PGg\PGg} < 120\GeV$ and $130 < m_{\PGg\PGg} < 180\GeV$.

\section{Event categorization}
Two BDT classifiers are constructed using the \textsc{XGBoost} package~\cite{chen2016xgboost}. The first, BDT1, is trained using events from simulated samples of the \cH signal and the \ggH background, which accounts for more than 60\% of the resonant background. The second, BDT2, is trained using events from simulated samples of the \cH signal and the continuous background.
The BDTs are constructed to have an output between 0 and 1.

\begin{figure}[!htb]
  \centering
  \includegraphics[width=0.49\textwidth]{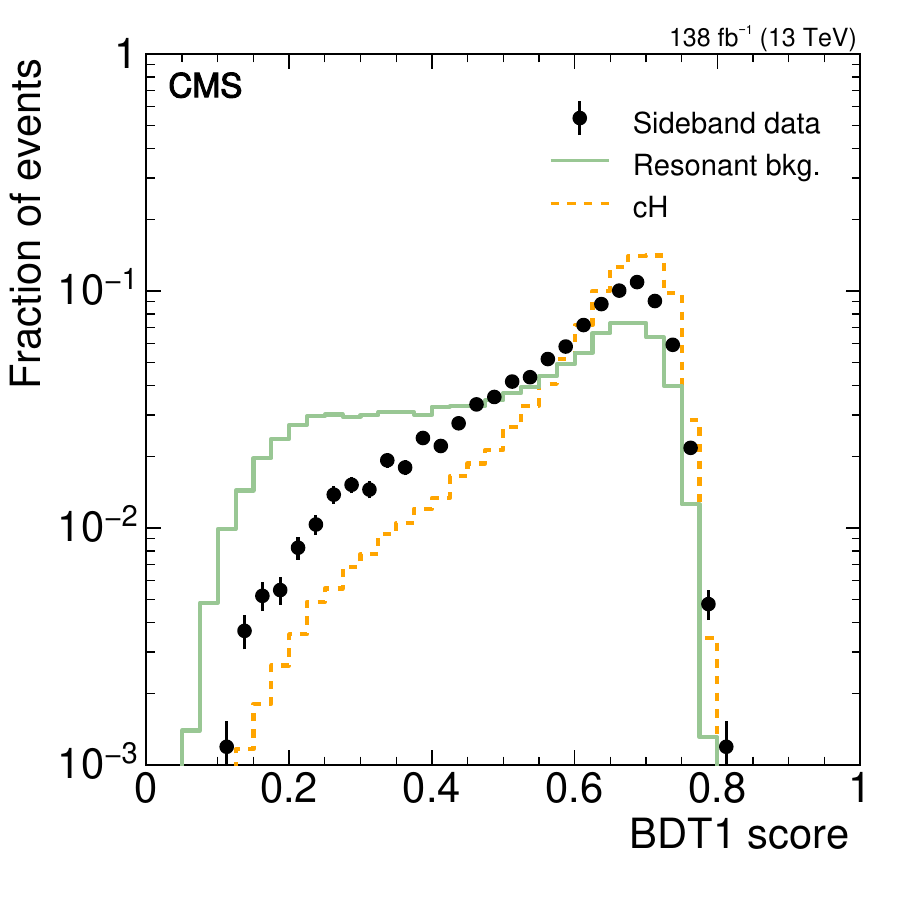}
  \includegraphics[width=0.49\textwidth]{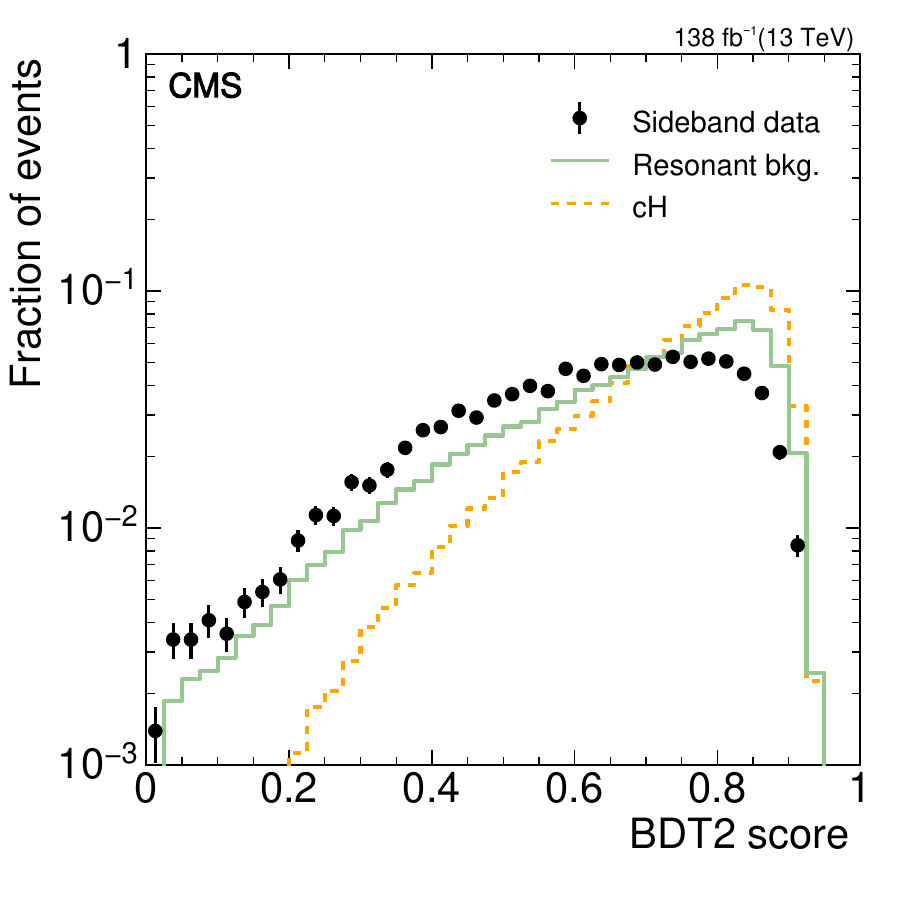}
  \caption{Distributions of BDT1 (\cmsLeft) and BDT2 outputs (\cmsRight) for simulated \cH signal and resonant background events, as well as sideband data (representing the continuous background) events. The BDT1 score discriminates the \cH signal from the resonant background. The BDT2 score discriminates the \cH signal from the continuous background. The plots are indicative of the shapes of the distributions and should not be interpreted as a comparison of their magnitudes. The error bars correspond to the statistical uncertainties.}
  \label{fig:BDTscore}
\end{figure}

For the training of both BDTs, the input features are $\pt/m_{\PGg\PGg}$, $\eta$, and photon ID~\cite{Sirunyan:2018ouh} score of the two leading photons, the $\pt$ and $\eta$ of the leading jet, the difference in $\phi$ (azimuthal angle) between the leading jet and each of the two photons, and the jet multiplicity.
As no simulation is available to reliably model events with high jet multiplicity, the jet multiplicity feature is a categorical variable with only three possible values: 1, 2, and more than 2.
The BDT input features are chosen based on their capacity to separate signal from background.
The BDT trainings avoid using \PQc tagging information, so that the event yield ratio between \ggH{+}\PQc jet and \ggH{+}(\PQu, \PQd, \PQs, \Pg) jet is not strongly correlated with the BDT outputs and hence the heavy flavor (HF) modeling uncertainty can be applied on the overall normalization of \ggH.
The use of tagging information in training classifiers can be further explored in future analyses.
Distributions of the outputs of the two BDTs for simulated \cH signal and resonant background, as well as sideband data (as defined above, representing the continuous background) events passing the selection described in Section~\ref{sec:selection}, are shown in Fig.~\ref{fig:BDTscore}.
The plots are indicative of the shapes of the distributions,
showing that BDT1 separates \cH from \ggH production and BDT2 separates \cH production from the continuous background.
The plots should not be interpreted as a comparison of their magnitudes.
The BDT1 plot shows the classifier is separating \cH and resonant background, and the BDT1 plots 

\begin{figure}[!htb]
  \centering
    \includegraphics[width=0.49\textwidth]{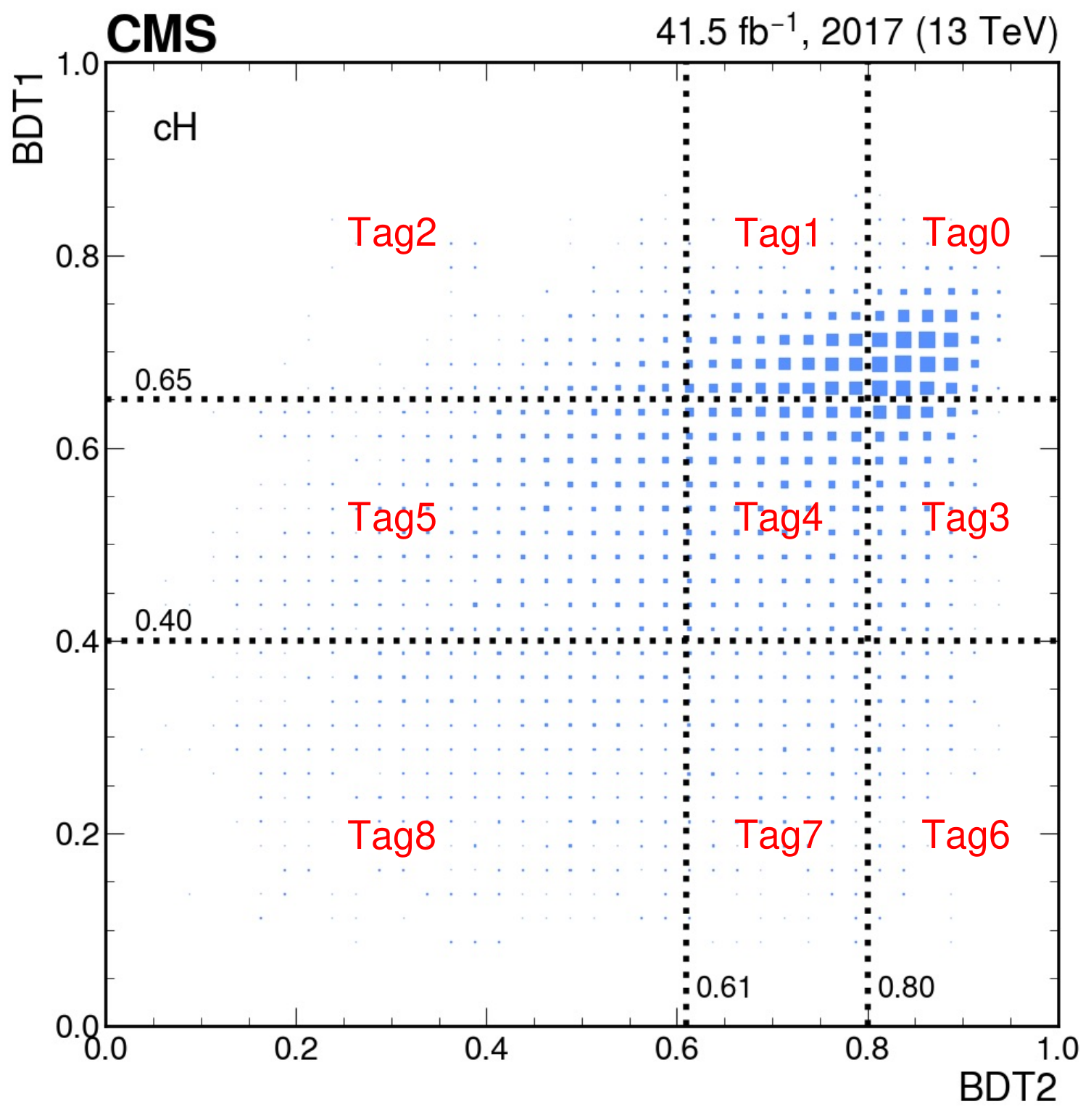} 
    \includegraphics[width=0.49\textwidth]{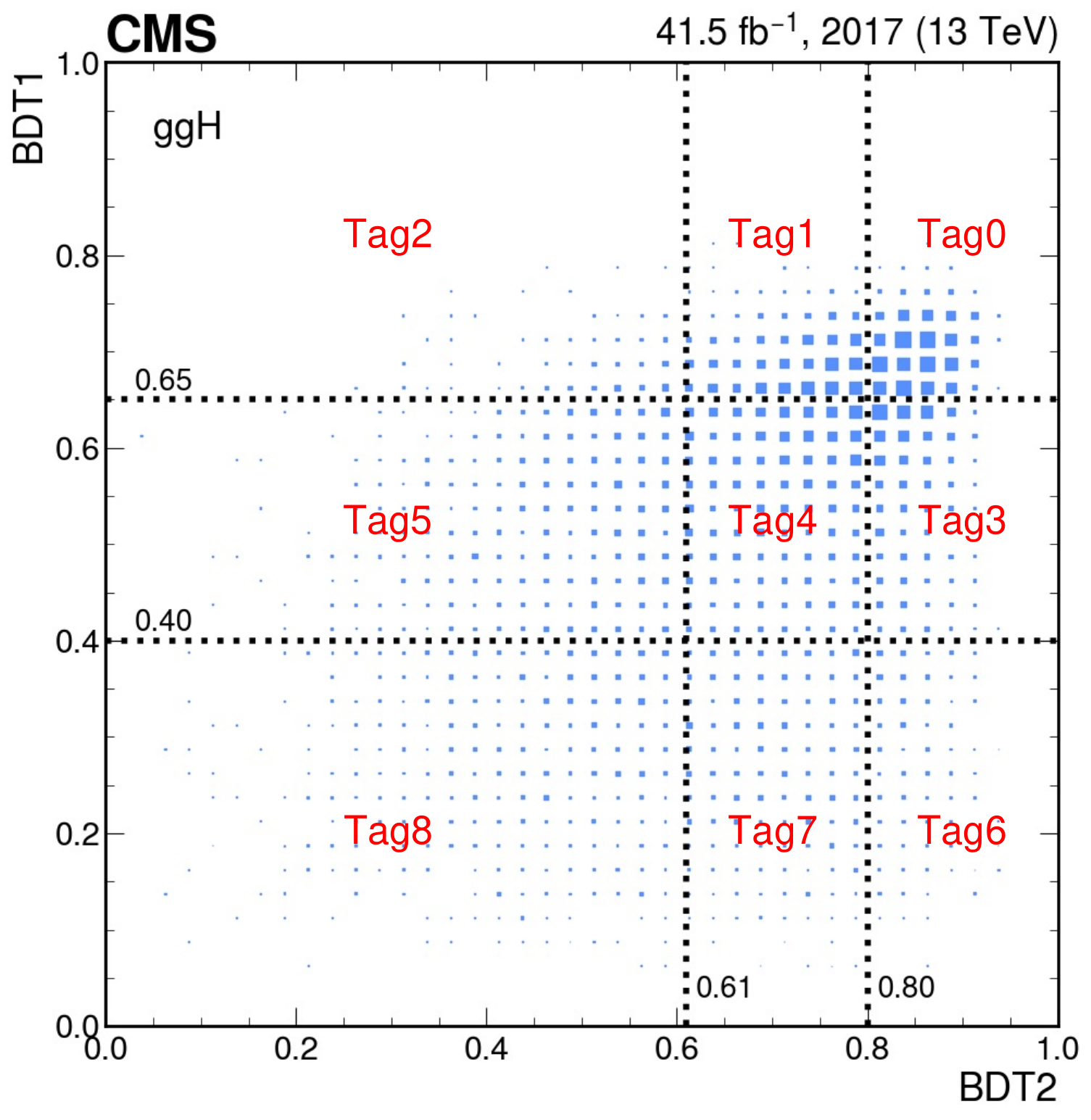} \\
    \includegraphics[width=0.49\textwidth]{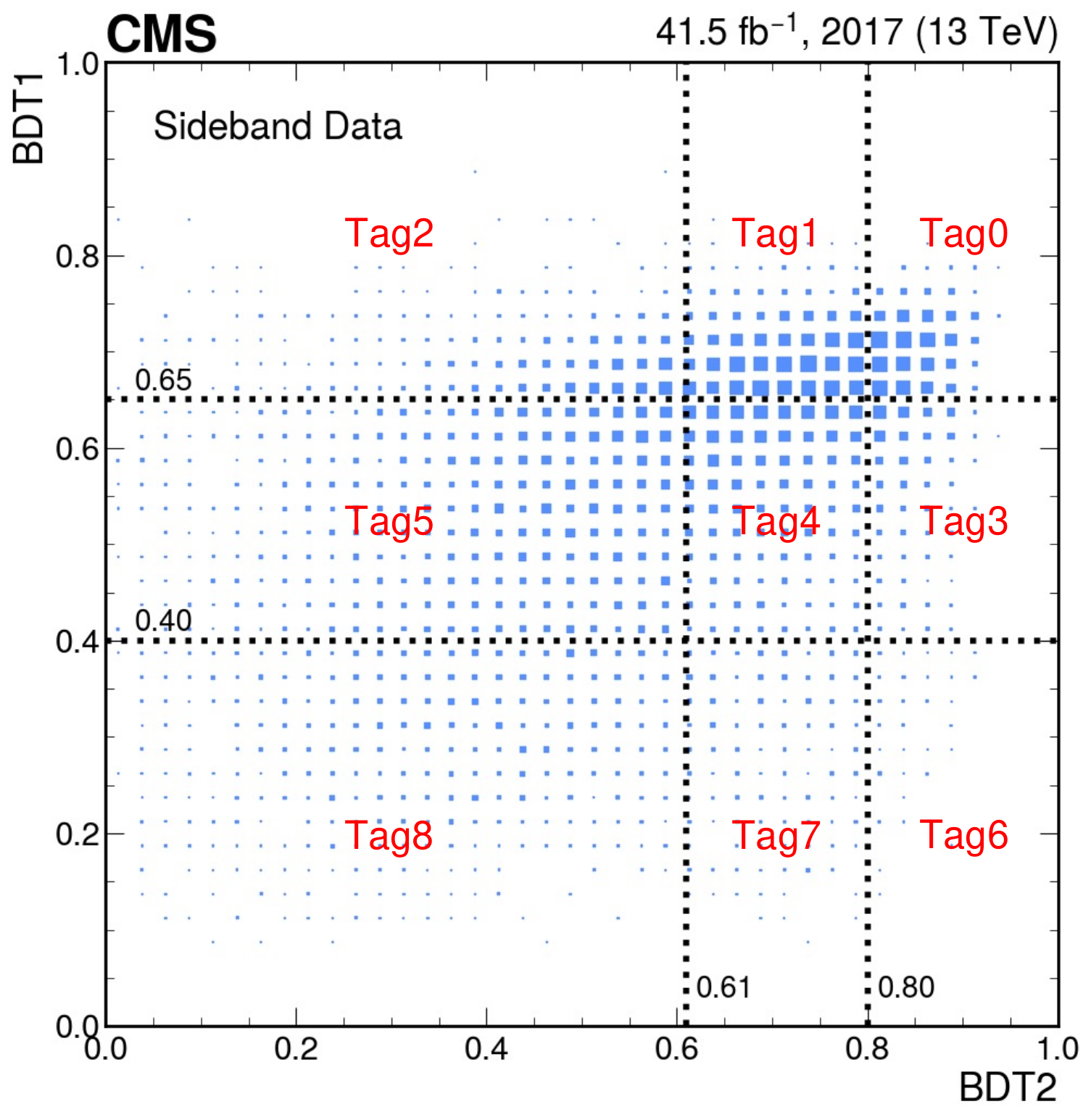} \\
  \caption{Two-dimensional distributions of BDT1 output ($y$ axis) and BDT2 output (x axis) for \cH (upper left), \ggH (upper right), and sideband data (lower) in 2017 event categories. The BDT1 score discriminates the \cH signal from the resonant background while the BDT2 score discriminates the \cH signal from the continuous background. Event category boundaries on BDT scores are illustrated as dashed lines.}
  \label{fig:categorization}
\end{figure}

A two-dimensional classification is created to separate the \cH signal from the main backgrounds.
First, events are split into three classes according to the output of BDT1.
The class boundaries are optimized to enhance the separation between the \cH and \ggH processes.
The optimization is performed using an approximate formula based on event yields.
Secondly, in each class, events are split into three categories according to the output of BDT2.
The category boundaries are optimized to maximize the expected sensitivity of the analysis to the \cH process.
For simplicity, the same boundaries are adopted for the three classes, whose sensitivity is close to the case where boundaries in the three classes are optimized independently. 
The nine event categories are created for each of the three data-taking years, resulting in 27 categories in total.
Two-dimensional distributions of BDT1 output and BDT2 output for \cH, \ggH, and sideband data in 2017 event categories are shown in Fig.~\ref{fig:categorization}.
The boundaries of the nine event categories Tag0--8 are also shown.

\section{Signal and background modeling}

\begin{figure}[!htb]
  \centering
  \includegraphics[width=0.49\textwidth]{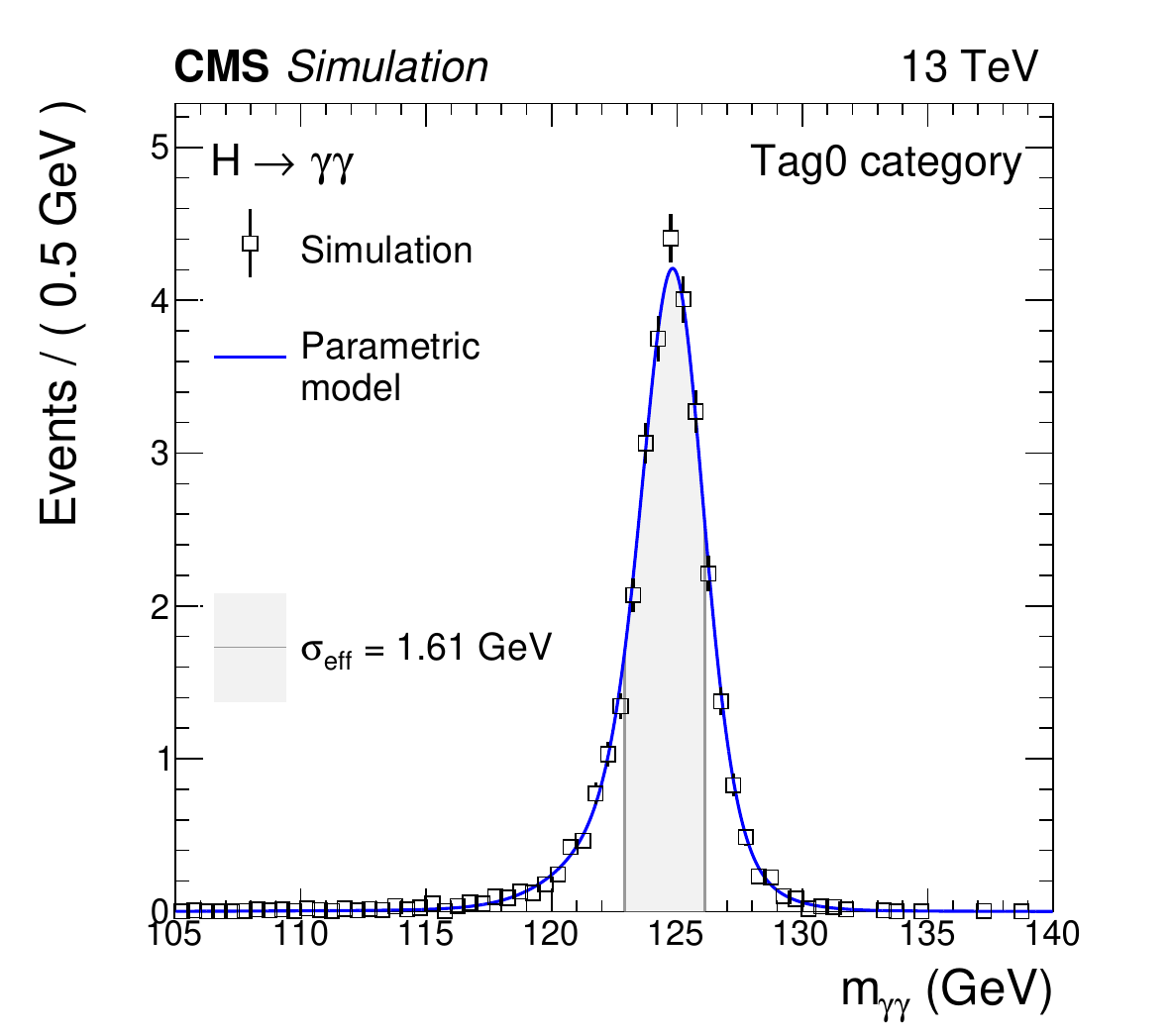}
  \includegraphics[width=0.49\textwidth]{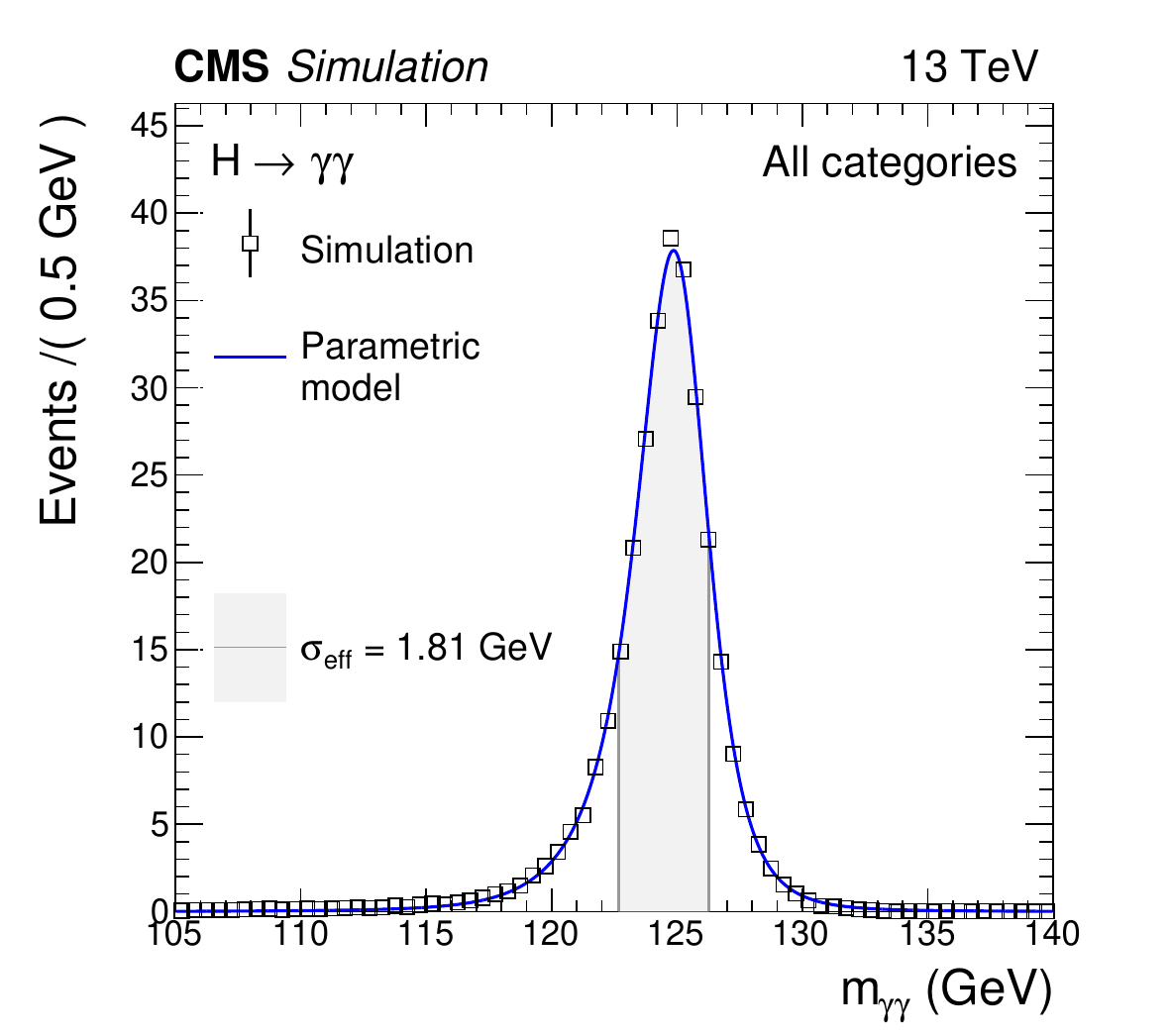}
  \caption{The shape of the parametric signal model for the analysis category with the tightest requirement in both BDT1 and BDT2 scores (\cmsLeft) and for the sum of all analysis categories (\cmsRight). The open squares represent simulated events and the blue line represents the corresponding model. Also shown is the $\sigma_{\text{eff}}$ value (half the width of the narrowest interval containing 68.3\% of the diphoton mass distribution) in the grey shaded area.}
  \label{fig:smodel}
\end{figure}

The diphoton mass distributions of the \cH signal and the other Higgs boson processes are modeled using sums of multiple Gaussian functions,
which are parametrized from simulation.
The Higgs boson mass is fixed to 125.38\GeV, which is the best CMS measurement combining the diphoton and four-lepton channels~\cite{CMS:2020xrn}. 
An F-test~\cite{fTest} is performed to determine the optimal number of Gaussian functions for the signal model.
In a given event category, a separate function is constructed for each Higgs boson process and the final signal model is obtained by summing the individual functions for all Higgs boson processes~\cite{Sirunyan:2018ouh}.
Figure~\ref{fig:smodel} shows signal models for the analysis category with the tightest requirement in both BDT1 and BDT2 scores (left) and for the sum of all analysis categories (right).

The continuous background mass distributions are modeled using data with the discrete profiling method~\cite{Dauncey:2014xga}, which treats the choice of the analytic function as a discrete nuisance parameter in the likelihood fit.
Several families of functions are considered, including exponential functions, Bernstein polynomials, Laurent series, and power law functions.
For each family, an F-test~\cite{fTest} using the data sidebands ($100 < m_{\PGg\PGg} < 120\GeV$ and $130 < m_{\PGg\PGg} < 180\GeV$) is performed to determine
the maximum number of degrees of freedom, while the minimum number of degrees of freedom is determined based on goodness-of-fit criterion in the data sidebands.
A signal injection test is also performed, which shows no significant bias of our background modeling.

\section{Systematic uncertainties}
\label{sec:sys}

The uncertainties in the continuous background estimation are accounted for by using the discrete profiling method.
Shape uncertainties in the mass distributions of the \cH signal and the other resonant background processes related to the photon energy scale and resolution and the assumed Higgs boson mass are considered as well.
They are incorporated in the fit model as nuisance parameters that affect the mean and width of the respective Gaussian functions.

The dominant theoretical uncertainties in the yields of the Higgs boson processes across the event categories are due to the choice of the flavor scheme for \cH and \bH production and the HF modeling for \ggH production. The typical size of the flavor scheme uncertainty is 25--30\%. For the HF modeling, we assume the theoretical uncertainty found in \ggH{+}\bbbar~\cite{Manzoni:2023qaf} and therefore apply a 50\% uncertainty in the inclusive \ggH yield, larger than the current uncertainty in the inclusive \ggH production. Future theoretical and experimental studies may reduce the size of this modeling uncertainty.
The uncertainties in the renormalization and factorization scales, the \alpS value, PDFs, underlying event and parton shower modeling, and the $\PH \to \PGg\PGg$ branching fraction are also considered. Their typical sizes are in the range of 1--20\%.
The theoretical uncertainties are correlated between the years.
The main experimental uncertainties are related to \PQc quark and photon identification as well as the jet and photon energy scales and resolutions. Their typical sizes are in the 2--10\% range.
Other experimental uncertainties, including those in the trigger efficiency, the diphoton preselection, and the pileup jet identification are less than 2\%.
Most of the experimental uncertainties are not correlated between the years.
The integrated luminosities for the 2016, 2017, and 2018 data-taking years have 1.2--2.5\% individual uncertainties~\cite{CMS-LUM-17-003,CMS-PAS-LUM-17-004,CMS-PAS-LUM-18-002}, while the overall uncertainty for the 2016--2018 period is 1.6\%.

\section{Results}

The following results have been obtained using the CMS statistical analysis tool \textsc{Combine}~\cite{CAT-23-001}.
To extract the signal strength $\mu_{\cH}$ of the \cH process, defined as the ratio of the measured yield to the one predicted by the SM,
we perform a simultaneous maximum likelihood fit to the $m_{\PGg\PGg}$ distributions in the 27 event categories.
In the likelihood fit, the yields of all resonant Higgs boson production modes assume their SM predictions and are allowed to vary within their uncertainties, while the normalization and shape parameters of the continuous background in each category are unconstrained. All systematic uncertainties are included as nuisance parameters in the profile likelihood ratio~\cite{CMS-NOTE-2011-005,Cowan:2010js}.

\begin{figure}[!htb]
  \centering
  \includegraphics[width=0.7\textwidth]{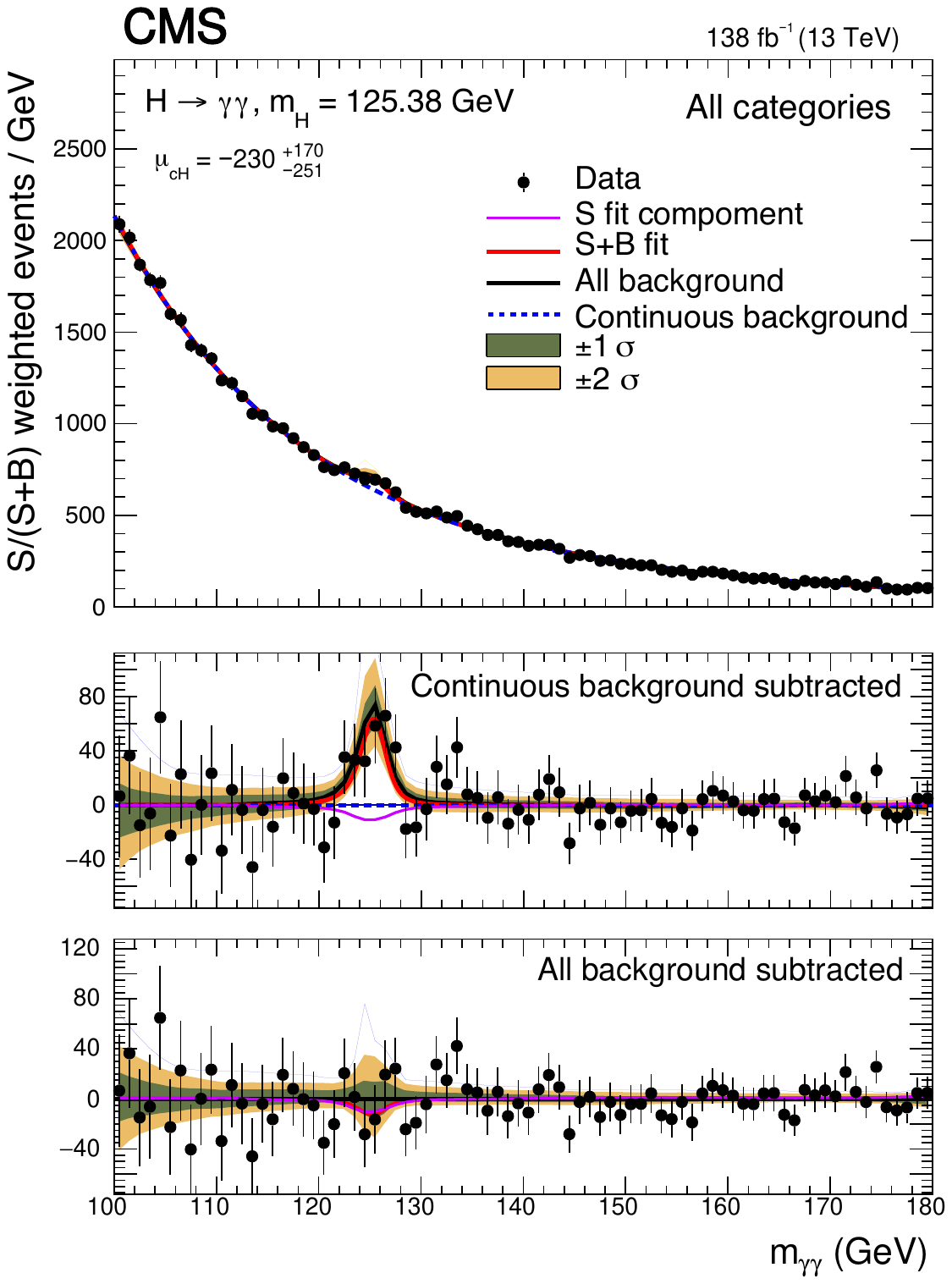}
  \caption{Invariant mass distribution of the selected events in all categories. For each category, the events are scaled by $\mathrm{S}/(\mathrm{S}+\mathrm{B})$, where S (B) is the number of expected signal (background) events in the smallest mass window containing 68.3\% of the expected signal events.
  Curves for the signal + background fit (red), separately showing the resonant (black) and continuous (blue) backgrounds, as well as bands representing the 68.3\% and 95.5\% \CL intervals in the background estimation, are overlaid.
  The middle (lower) panel shows the $m_{\PGg\PGg}$ distribution after subtracting only the continuous background (subtracting all background components) and overlaying the curve for the fitted signal (purple).
  }
  \label{fig:result_mass_postfit}
\end{figure}

The invariant mass distribution of the selected events in all categories is shown in Fig.~\ref{fig:result_mass_postfit}.
For each category, the events are scaled by $\mathrm{S}/(\mathrm{S}+\mathrm{B})$, where S (B) is the number of expected signal (background) events in the smallest mass window containing 68.3\% of the expected signal events.
Curves for the signal + background fit, separately showing the resonant and continuous backgrounds, as well as bands representing the 68.3\% and 95.5\% \CL intervals in the background estimation, are overlaid.
The resonant background visible in the figure is dominated by \ggH production, where the leading jet consists of 25\% \PQc jet, 19\% \PQb jet and 57\% (\PQu, \PQd, \PQs, \Pg) jet after the $\text{CvsL} > 0.25$ selection.
Table~\ref{tab:yield_summary} shows the number of expected signal, resonant background and continuous background events in the mass window $[122.88,127.88]\GeV$ around the assumed Higgs boson mass of 125.38\GeV~\cite{CMS:2020xrn} for all categories. The size of the mass window is about twice the diphoton invariant mass resolution.

\begin{table}[!htbp]
    \centering
    \topcaption{Number of expected signal \cH ($\PH \to \PGg\PGg$), resonant background and continuous background events, as well as the resulting signal-over-background ratio (S/B) in the diphoton mass window $[122.88, 127.88]\GeV$ for all categories. For each category, the event yields for the three years are summed. The fraction of different production processes contributing to the resonant background (\ggH, \ttH, VBF, \VH, and \bH) is also reported.
    }
      \begin{tabular}{l l rrrrr r r r}
        Category & \multicolumn{1}{c}{Signal} & \multicolumn{6}{c}{Resonant background} & \multicolumn{1}{c}{Continuous} & \multicolumn{1}{c}{S/B} \\
        & \multicolumn{1}{c}{\cH} & \multicolumn{1}{c}{\ggH} & \multicolumn{1}{c}{\ttH} & \multicolumn{1}{c}{VBF} & \multicolumn{1}{c}{\VH} & \multicolumn{1}{c}{\bH} & \multicolumn{1}{c}{Total} &  bkg. ($\times 10^3$) & ($\times 10^{-5}$) \\
        \hline
        Tag0 & 0.013 & 84\% & $<$0.1\% & 5.3\% & 3.4\% & 7.5\% & 2.4 & 0.50 &2.6 \\
        Tag1 & 0.016 & 79\% & 0.33\% & 7.3\% & 6.3\% & 7.3\% & 3.3 & 1.5 &1.0 \\
        Tag2 & 0.0072 & 72\% & 4.0\% & 8.3\% & 9.1\% & 6.4\% & 1.8 & 7.4 & 0.10\\
        Tag3 & 0.0034 & 72\% & $<$0.1\% & 16\% & 5.9\% & 5.6\% & 1.3 & 0.17 & 2.0\\
        Tag4 & 0.0087 & 68\% & 1.2\% & 16\% & 9.9\% & 4.9\% & 3.5 & 0.96 &0.90\\
        Tag5 & 0.0094 & 54\% & 15\% & 15\% & 14\% & 3.6\% & 5.1 & 9.9 &0.10\\
        Tag6 & 0.00029 & 42\% & 1.9\% & 43\% & 12\% & 1.5\% & 0.52 & 0.019 &1.5\\
        Tag7 & 0.00095 & 43\% & 14\% & 25\% & 17\% & 1.3\% & 1.8 & 0.16 &0.59\\
        Tag8 & 0.0017 & 36\% & 32\% & 15\% & 17\% & 1.1\% & 3.3 & 1.9 &0.087\\[\cmsTabSkip]
        All & 0.060 & 61\% & 9.4\% & 14\% & 11\% & 4.5\% & 23 & 23 &0.27\\
      \end{tabular}
    \label{tab:yield_summary}
\end{table}

The observed data are compatible with the background-only hypothesis.
The observed $\mu_{\cH}$ is evaluated to be $-230^{+170}_{-251}$.
Using asymptotic distributions for the test statistic based on the profile likelihood ratio~\cite{CMS-NOTE-2011-005,Cowan:2010js},
an observed (expected) upper limit at 95\% \CL on $\mu_{\cH}$ is set at 243\,(355) times the SM prediction.

Furthermore, a specific fit is performed to set constraints on \kappac.
Taking the ``flat direction'' approach introduced in Ref.~\cite{Coyle:2019hvs}, the modifier of the other Higgs coupling strengths, $\kappa_\PH$, depends on \kappac and the branching fraction of the $\PH \to \ccbar$ decay predicted by the SM, $\mathcal{B}^\mathrm{SM}_{\ccbar}$, so that the signal strengths of non-\cH Higgs boson processes in the diphoton decay channel are always unity:
\begin{equation}
\kappa_\PH^{2}  = \frac{(1 -  \mathcal{B}^\mathrm{SM}_{\ccbar})}{2} + \frac{\sqrt{(1 -  \mathcal{B}^\mathrm{SM}_{\ccbar})^{2} + 4  \mathcal{B}^\mathrm{SM}_{\ccbar} \kappac^{2}}}{2}.
\end{equation}
The signal strength of the \cH process $\mu_{\cH}$ can, however, be modified by \kappac through the production cross section as well as partial and total decay widths:
\begin{equation}
\mu_{\cH}  = \frac{\kappac^{2} \kappa_\PH^{2}}{(1 -  \mathcal{B}^\mathrm{SM}_{\ccbar})\kappa_\PH^{2} +  \mathcal{B}^\mathrm{SM}_{\ccbar} \kappac^{2}}.
\end{equation}
At 95\% \CL, the observed (expected) constraint is $\abs{\kappac} < 38.1\,(72.5)$.
The confidence intervals presented in this analysis are based on an asymptotic approximation.
As demonstrated in Ref.~\cite{CMS:2025ugn},
the coverage achieved using this asymptotic method aligns reasonably well with the intended coverage across all tested parameters,
including scenarios similar to this analysis.

The sensitivity of this analysis is limited by the statistical uncertainty in the data, and the systematic uncertainties in the choice of flavor scheme, factorization and renormalization scales for \cH production, and the HF modeling of \ggH production.
Table~\ref{tab:syst_summary} shows the impacts of several uncertainty groups divided by the total uncertainty in the signal strength measurement.

\begin{table}[!htbp]
  \centering
  \topcaption{Impacts of several uncertainty groups divided by the total uncertainty in the signal strength measurement. The systematic uncertainties are described in Sec.~\ref{sec:sys}.
  }
  \begin{tabular}{ll}
    Uncertainty group & Fraction of total impact \\
    \hline
    Statistical     & 66\% \\
    Theoretical in the \cH signal     & 38\% \\
    Theoretical in the resonant background     & 59\% \\
    Experimental in the yields     & 27\% \\
    Experimental in the mass shapes     & $<$1\% \\
    Integrated luminosity    & $<$1\% \\
  \end{tabular}
  \label{tab:syst_summary}
\end{table}

As a cross-check of the resonant background normalization, we perform another maximum likelihood fit floating the inclusive signal strength of Higgs boson production (including \cH, \ggH, and the other Higgs boson production modes) in all analysis categories.
The observed (expected) inclusive signal strength is evaluated to be $0.8^{+0.8}_{-0.4}$ ($1.0^{+0.7}_{-0.4}$), also compatible with the SM prediction.

Tabulated results are provided in the HEPData record for this analysis~\cite{hepdata}.

\section{Summary}
We have presented the first search for a \cH signal sensitive to the coupling of the charm quark (\PQc) to the Higgs boson (\PH) in the associated production of at least one charm quark with a Higgs boson in proton-proton collisions at a center-of-mass energy of 13 TeV.
Assuming the signal strengths of non-\cH Higgs production processes in the diphoton decay channel to be as predicted by the standard model (SM), the observed and expected upper limits at 95\% confidence level on the \cH signal strength are 243 and 355 times the SM predictions, respectively.
This search also provides sensitivity to the Yukawa coupling between the Higgs boson and the charm quark.
Under the same assumption, the observed (expected) allowed interval on the Higgs boson to charm quark coupling modifier \kappac is $\abs{\kappac} < 38.1$ ($\abs{\kappac} < 72.5$) at 95\% confidence level.
The analysis is presently limited by statistical and theoretical uncertainties. Improvements can be achieved by incorporating complementary Higgs boson decay channels and, as larger data sets become available, implementing more refined selection criteria and categorization strategies, such as discriminating charm and bottom quark jets and constructing classifiers distinguishing between more Higgs boson production modes.

\begin{acknowledgments}
We thank M.~Zaro and D.~Pagani for their help in the calculation and generation of the signal process.
  
We congratulate our colleagues in the CERN accelerator departments for the excellent performance of the LHC and thank the technical and administrative staffs at CERN and at other CMS institutes for their contributions to the success of the CMS effort. In addition, we gratefully acknowledge the computing centers and personnel of the Worldwide LHC Computing Grid and other centers for delivering so effectively the computing infrastructure essential to our analyses. Finally, we acknowledge the enduring support for the construction and operation of the LHC, the CMS detector, and the supporting computing infrastructure provided by the following funding agencies: SC (Armenia), BMBWF and FWF (Austria); FNRS and FWO (Belgium); CNPq, CAPES, FAPERJ, FAPERGS, and FAPESP (Brazil); MES and BNSF (Bulgaria); CERN; CAS, MoST, and NSFC (China); MINCIENCIAS (Colombia); MSES and CSF (Croatia); RIF (Cyprus); SENESCYT (Ecuador); ERC PRG, RVTT3 and MoER TK202 (Estonia); Academy of Finland, MEC, and HIP (Finland); CEA and CNRS/IN2P3 (France); SRNSF (Georgia); BMBF, DFG, and HGF (Germany); GSRI (Greece); NKFIH (Hungary); DAE and DST (India); IPM (Iran); SFI (Ireland); INFN (Italy); MSIP and NRF (Republic of Korea); MES (Latvia); LMTLT (Lithuania); MOE and UM (Malaysia); BUAP, CINVESTAV, CONACYT, LNS, SEP, and UASLP-FAI (Mexico); MOS (Montenegro); MBIE (New Zealand); PAEC (Pakistan); MES and NSC (Poland); FCT (Portugal); MESTD (Serbia); MICIU/AEI and PCTI (Spain); MOSTR (Sri Lanka); Swiss Funding Agencies (Switzerland); MST (Taipei); MHESI and NSTDA (Thailand); TUBITAK and TENMAK (Turkey); NASU (Ukraine); STFC (United Kingdom); DOE and NSF (USA).

\hyphenation{Rachada-pisek} Individuals have received support from the Marie-Curie program and the European Research Council and Horizon 2020 Grant, contract Nos.\ 675440, 724704, 752730, 758316, 765710, 824093, 101115353, 101002207, and COST Action CA16108 (European Union); the Leventis Foundation; the Alfred P.\ Sloan Foundation; the Alexander von Humboldt Foundation; the Science Committee, project no. 22rl-037 (Armenia); the Fonds pour la Formation \`a la Recherche dans l'Industrie et dans l'Agriculture (FRIA-Belgium); the Beijing Municipal Science \& Technology Commission, No. Z191100007219010 and Fundamental Research Funds for the Central Universities (China); the Ministry of Education, Youth and Sports (MEYS) of the Czech Republic; the Shota Rustaveli National Science Foundation, grant FR-22-985 (Georgia); the Deutsche Forschungsgemeinschaft (DFG), among others, under Germany's Excellence Strategy -- EXC 2121 ``Quantum Universe" -- 390833306, and under project number 400140256 - GRK2497; the Hellenic Foundation for Research and Innovation (HFRI), Project Number 2288 (Greece); the Hungarian Academy of Sciences, the New National Excellence Program - \'UNKP, the NKFIH research grants K 131991, K 133046, K 138136, K 143460, K 143477, K 146913, K 146914, K 147048, 2020-2.2.1-ED-2021-00181, TKP2021-NKTA-64, and 2021-4.1.2-NEMZ\_KI-2024-00036 (Hungary); the Council of Science and Industrial Research, India; ICSC -- National Research Center for High Performance Computing, Big Data and Quantum Computing and FAIR -- Future Artificial Intelligence Research, funded by the NextGenerationEU program (Italy); the Latvian Council of Science; the Ministry of Education and Science, project no. 2022/WK/14, and the National Science Center, contracts Opus 2021/41/B/ST2/01369 and 2021/43/B/ST2/01552 (Poland); the Funda\c{c}\~ao para a Ci\^encia e a Tecnologia, grant CEECIND/01334/2018 (Portugal); the National Priorities Research Program by Qatar National Research Fund; MICIU/AEI/10.13039/501100011033, ERDF/EU, "European Union NextGenerationEU/PRTR", and Programa Severo Ochoa del Principado de Asturias (Spain); the Chulalongkorn Academic into Its 2nd Century Project Advancement Project, and the National Science, Research and Innovation Fund via the Program Management Unit for Human Resources \& Institutional Development, Research and Innovation, grant B39G670016 (Thailand); the Kavli Foundation; the Nvidia Corporation; the SuperMicro Corporation; the Welch Foundation, contract C-1845; and the Weston Havens Foundation (USA).
\end{acknowledgments}

\bibliography{auto_generated} 

\cleardoublepage \appendix\section{The CMS Collaboration \label{app:collab}}\begin{sloppypar}\hyphenpenalty=5000\widowpenalty=500\clubpenalty=5000
\cmsinstitute{Yerevan Physics Institute, Yerevan, Armenia}
{\tolerance=6000
V.~Chekhovsky, A.~Hayrapetyan, V.~Makarenko\cmsorcid{0000-0002-8406-8605}, A.~Tumasyan\cmsAuthorMark{1}\cmsorcid{0009-0000-0684-6742}
\par}
\cmsinstitute{Institut f\"{u}r Hochenergiephysik, Vienna, Austria}
{\tolerance=6000
W.~Adam\cmsorcid{0000-0001-9099-4341}, J.W.~Andrejkovic, L.~Benato\cmsorcid{0000-0001-5135-7489}, T.~Bergauer\cmsorcid{0000-0002-5786-0293}, S.~Chatterjee\cmsorcid{0000-0003-2660-0349}, K.~Damanakis\cmsorcid{0000-0001-5389-2872}, M.~Dragicevic\cmsorcid{0000-0003-1967-6783}, P.S.~Hussain\cmsorcid{0000-0002-4825-5278}, M.~Jeitler\cmsAuthorMark{2}\cmsorcid{0000-0002-5141-9560}, N.~Krammer\cmsorcid{0000-0002-0548-0985}, A.~Li\cmsorcid{0000-0002-4547-116X}, D.~Liko\cmsorcid{0000-0002-3380-473X}, I.~Mikulec\cmsorcid{0000-0003-0385-2746}, J.~Schieck\cmsAuthorMark{2}\cmsorcid{0000-0002-1058-8093}, R.~Sch\"{o}fbeck\cmsAuthorMark{2}\cmsorcid{0000-0002-2332-8784}, D.~Schwarz\cmsorcid{0000-0002-3821-7331}, M.~Sonawane\cmsorcid{0000-0003-0510-7010}, W.~Waltenberger\cmsorcid{0000-0002-6215-7228}, C.-E.~Wulz\cmsAuthorMark{2}\cmsorcid{0000-0001-9226-5812}
% ...
\par}
\vskip\cmsinstskip
\dag:~Deceased\\
$^{1}$Also at Yerevan State University, Yerevan, Armenia\\
$^{2}$Also at TU Wien, Vienna, Austria\\
% ...
\end{sloppypar}
\end{document}